\documentclass[preprint,12pt]{elsarticle}

\usepackage{amssymb}
\usepackage{multirow}
\usepackage{wrapfig}

\usepackage{tabularx}
\usepackage{booktabs}

\usepackage[utf8]{inputenc}
\usepackage{amsmath}
\usepackage{amsfonts}
\usepackage{epsfig}
\usepackage{psfrag}
\usepackage{pstricks}
\usepackage{algorithm}
\usepackage{graphicx}
\graphicspath{{./Figures/}}

\journal{Elsevier}

\begin{document}

\begin{frontmatter}



\title{Soap Film-inspired Subdivisional Lattice Structure Construction}

\author[]{Guoyue Luo}
\author[]{Qiang Zou\corref{cor}}\ead{qiangzou@cad.zju.edu.cn}

\cortext[cor]{Corresponding author.}
\address{State Key Laboratory of CAD$\&$CG, Zhejiang University, Hangzhou, 310058, China}

\begin{abstract}
Lattice structures, distinguished by their customizable geometries at the microscale and outstanding mechanical performance, have found widespread application across various industries. One fundamental process in their design and manufacturing is constructing boundary representation (B-rep) models, which are essential for running advanced applications like simulation, optimization, and process planning. However, this construction process presents significant challenges due to the high complexity of lattice structures, particularly in generating nodal shapes where robustness and smoothness issues can arise from the complex intersections between struts. To address these challenges, this paper proposes a novel approach for lattice structure construction by cutting struts and filling void regions with subdivisional nodal shapes. Inspired by soap films, the method generates smooth, shape-preserving control meshes using Laplacian fairing and subdivides them through the point-normal Loop (PN-Loop) subdivision scheme to obtain subdivisional nodal shapes. The proposed method ensures robust model construction with reduced shape deviations, enhanced surface fairness, and smooth transitions between subdivisional nodal shapes and retained struts. The effectiveness of the method has been demonstrated by a series of examples and comparisons. The code will be open-sourced upon publication.
\end{abstract}

\begin{keyword}
Computer-Aided Design \sep Lattice Structures \sep Geometric Modeling \sep Boundary Evaluation \sep Soap Film \sep Subdivision Surfaces
\end{keyword}

\end{frontmatter}


\section{Introduction}
With the advances in additive manufacturing, lattice structures have seen increasing applications in fields such as aerospace, automotive, and energy industries~\cite{2019_Nazir_cellular-structures-review_IJAMT}. These structures with interconnected struts at the microscopic scale can exhibit a range of various superior mechanical properties, e.g., lightweight, high strength, and multifunctional~\cite{2021_Obadimu_lattice_compressieve-behavior}, by precisely controlling the microscale geometries. This customizability of lattice structures allows for the tailored design of components that meet specific performance requirements.

Extensive research has been conducted to investigate the mechanical behavior of lattice structures and establish the ``shape-property-process" relationship through both theoretical and empirical approaches~\cite{wu2019design}. However, as recently noted by Verma et al.~\cite{2020_convexHull} and others~\cite{2020_CHoCC,2018_subdivision_2Dsurface,2005_HongqingWang_2Dsurface}, researchers have consistently encountered challenges in constructing boundary representation (B-rep) models for lattice structures while developing the relationship. Additionally, B-rep models of lattice structures can provide greater flexibility in shape manipulation compared to implicit models. They can also enable native compatibility with modern computer-aided design, engineering analysis, and manufacturing software pipeline~\cite{tpms2step,2024_meta-meshing}. However, traditional solid modeling techniques in computer-aided design~\cite{zou2023variational,2019_push-pull-direct} often fail to construct lattice structures because they cannot efficiently handle a much larger number of geometric elements and the much complex way of how elements intersect with each other~\cite{2025_microstructure-survey}.

\begin{figure}[htbp]
\centering
\includegraphics[width=0.6\linewidth]{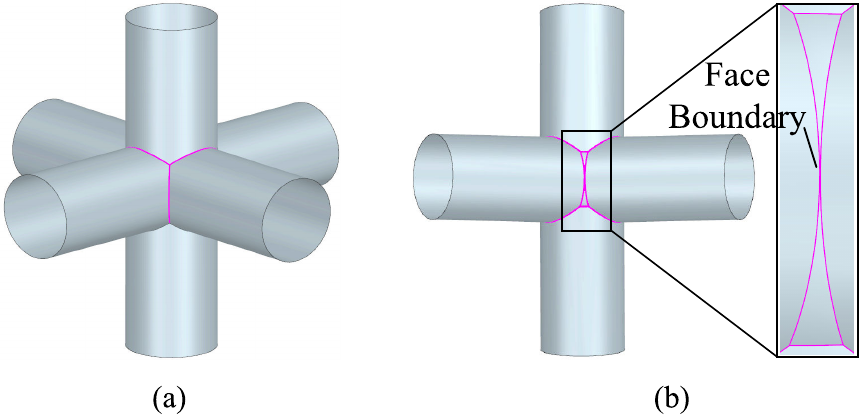}
\caption{Boolean intersections of multiple struts: (a) a regular case; (b) a singular example (tangency).}
\label{fig:intersection}
\end{figure}

The key challenge of lattice structure construction lies in the robust generation of nodal shapes. Conventionally, they are generated using Boolean operations, which compute the intersections of multiple struts (cylinders) to form their boundary shapes, as seen in Fig.~\ref{fig:intersection}a. However, these operations are notoriously unreliable~\cite{2020_convexHull,2021_1D-skeleton_Zouqiang_convolution-surface}. Although struts (i.e., cylinders) are seemingly simple to intersect, robustness issues still occur. As already highlighted by Hoffman~\cite{1989_Hoffmann_problems}, even for such simple geometries, singular tangencies (see Fig.~\ref{fig:intersection}b) can cause intersection failures. Furthermore, even if intersections are error-free, sharp edges can be produced, leading to stress concentration and, therefore, a significant performance decrease in practical engineering~\cite{2017_stress-relief}. Additionally, these sharp features can complicate the additive manufacturing process control and introduce difficulties in maintaining consistent printing quality~\cite{2019_AM-quality-sharp-corner}.

Recently, the local replacement has emerged as a solution to these problems, which replaces original nodal shapes (the shape resulting from the intersection of struts~\footnote{Usually, a sphere with the same radius as the struts is incorporated in Boolean intersections to prevent the occurrence of open nodal shapes, e.g, when only two struts intersect.}) with simpler ones~\cite{2020_CHoCC,2020_convexHull,2018_subdivision_2Dsurface,2023_2Dsurface_Xiong_Subdivisional-modelling}. Existing methods related to this topic may be classified into two categories. The first one is to replace original nodal shapes with simplified convex hulls~\cite{2020_CHoCC,2020_convexHull}. The second category approximates the original nodal shapes using subdivision surfaces~\cite{2018_subdivision_2Dsurface,2023_2Dsurface_Xiong_Subdivisional-modelling}. 

Such replacements can effectively avoid complex intersections and obtain relatively smoother transitions among struts. However, these convex hull nodes lack smoothness, both in their surfaces and in the transitions between them and the retained struts. Comparatively, subdivision surfaces can inherently guarantee smoothness through subdivision rules and allow for topology-free construction~\cite{2013_subDSurface_cad} that can handle the complex boundaries of the nodes after cutting. Nevertheless, the proposed subdivisional construction method~\cite{2018_subdivision_2Dsurface} is primarily developed for cubic cells, with the extension to other cell types being briefly discussed. Moreover, it does not consider overall surface fairness and the shape deviations between the subdivisional nodal shape and the original one.

In this paper, we attempt to construct subdivisional nodal shapes that exhibit enhanced surface fairness and reduced shape deviation, and ensure smooth transitions between the nodal shapes and the retained struts. Specifically, we cut struts to separate the replaceable portion of the lattice from the parts to be retained, with the void volume minimized to avoid large shape modification. These void regions are then filled using a soap film-inspired nodal shape construction method. The proposed method can effectively reduce shape deviations by optimizing the film geometry through simple line-cylinder intersections and improve surface fairness via cost-effective Laplacian fairing. Besides those, the proposed method derived the combined PN-Loop scheme from the generalized point-normal subdivision framework~\cite{2023_PN-subdivision}. This scheme can not only generate a smooth and fair surface but also achieve smooth transitions between nodal shapes and retained struts.

The following sections begin with a review of related work in Section~\ref{sec:related}, then method details in Section~\ref{sec:methods}, then validation of the method using a series of examples and comparisons in Section~\ref{sec:results}, and finally conclusions on the method’s advantages and limitations in Section~\ref{sec:conclusion}.

 \section{Related Work}
\label{sec:related}
In this section, we briefly discuss literature related to lattice structure modeling. The existing methods may be categorized as explicit and implicit. Implicit methods represent lattice structures using mathematical functions, whereas explicit methods evaluate the boundaries of lattice structures. The evaluation methods may include isosurface extraction, node simplification, and node approximation.

\subsection{Explicit Methods}
These kinds of methods directly store the boundary elements of lattice structures, such as vertices, edges, and faces~\cite{requicha1982solid}. This approach offers distinct advantages for downstream applications (e.g., engineering analysis~\cite{li2023xvoxel} and process planning~\cite{zou2013iso,zou2014iso,zou2021length}). Hence, various methods have been proposed to evaluate the boundaries, including isosurface extraction, node simplification, and node approximation.

\textit{Evaluation through isosurface extraction.} Direct modeling of lattice structures using Boolean operations encounters robustness issues, prompting the adoption of indirect methods for more reliable evaluation. A common approach involves representing lattice structures as implicit, which is then evaluated through various techniques. Among these, the marching cubes method~\cite{2021_Stromberg_MC-TPMS} and the marching tetrahedra method~\cite{2022_Ma_MT-TPMS} have been widely employed. However, it should be noted that the use of these evaluation methods may yield low-quality results with redundant facets produced, raising concerns about representational accuracy~\cite{2024_meta-meshing}. This limitation can become particularly problematic when modeling lattice structures with fine-scale geometries. Additionally, due to the inherent limitations of implicit models, fine-tuning nodal shapes to achieve smooth transitions between struts is challenging.

\textit{Evaluation through node simplification.} A recently emerging lattice structure evaluation strategy is to locally replace the original nodal shapes using simplified nodes. As involved in~\cite{2020_CHoCC,2020_convexHull}, the convex hull nodes are used as replacements for the original nodes of lattice structures. These convex hulls are then cohesively integrated with the retained struts to form the overall structure---combining into a solid B-rep model with valid topology~\cite{zou2022robust}. This simplification helps avoid complex surface-surface intersections, facilitating the robust construction of lattice structures. However, the surface of the convex hull is typically not smooth (e.g., it often contains creases), and smooth transition issues persist between the convex hulls and the retained struts. These factors can lead to a reduction in the mechanical performance of the lattice structure.

\textit{Evaluation through node approximation.} Unlike discrete convex hulls, another kind of replacement scheme constructs smooth subdivision surfaces to approximate the original nodal shapes~\cite{2018_subdivision_2Dsurface,2023_2Dsurface_Xiong_Subdivisional-modelling}. Subdivision surfaces, defined as piecewise parametric surfaces over meshes of arbitrary topology, combine the topological flexibility of meshes with the smoothness of parametric surfaces. This dual advantage makes them a valuable tool for modeling lattice structures. However, the existing approximation method proposed by Savio~\cite{2018_subdivision_2Dsurface} is primarily developed for specialized cells (e.g., cubic cells) and neglects critical factors such as shape deviations from the original nodal shapes and the overall surface fairness. These limitations are addressed in our proposed method.

\subsection{Implicit Methods}
Implicit methods describe geometric shapes using scalar functions, with isocontours defining their boundary surfaces. This approach simplifies Boolean operations such as unions and intersections through basic min/max comparisons, ensuring robustness and simplicity even for complex geometries. Given these advantages, various scalar functions, including signed distance fields~\cite{2020_1D-skeleton_Liang_VDF-conformal}, convolutional surfaces~\cite{2019_1D-skeleton_Tang_convolution-surface,2021_1D-skeleton_Zouqiang_convolution-surface}, implicit splines~\cite{2023_Elber_implicit-conformal_CAD}, and general mathematical formulations~\cite{2011_Pasko_FrepLattice,2013_4D-rep_Fryazinov_Frep_multi-scale,ding2021stl}, have been used for modeling intricate lattice structures. The practical effectiveness of implicit modeling is demonstrated by its implementation in commercial software packages like nTopology~\cite{nTopology} and IceSL~\cite{iceSL}.

Despite wide applications, these methods have limitations. A major challenge lies in accurately representing complex lattice structures, which feature intricate geometries and irregular topologies. These structures cannot be described by a single continuous mathematical formula; instead, they are typically represented using discrete implicit fields, which depend heavily on the field resolution. Low-resolution fields often fail to capture the fine-scale geometric details, while high-resolution fields demand significant computational resources and memory~\cite{1995_adaptive-MC}. Additionally, these implicit lattice structures generally need to be converted to explicit formats through isosurface extraction for additive manufacturing~\cite{tpms2step}. However, as aforementioned, this conversion process can lead to redundant facets.

\section{Methods}
\label{sec:methods}
\subsection{Soap Film-inspired Framework}
\begin{figure}[htbp]
\centering
\includegraphics[width=0.6\linewidth]{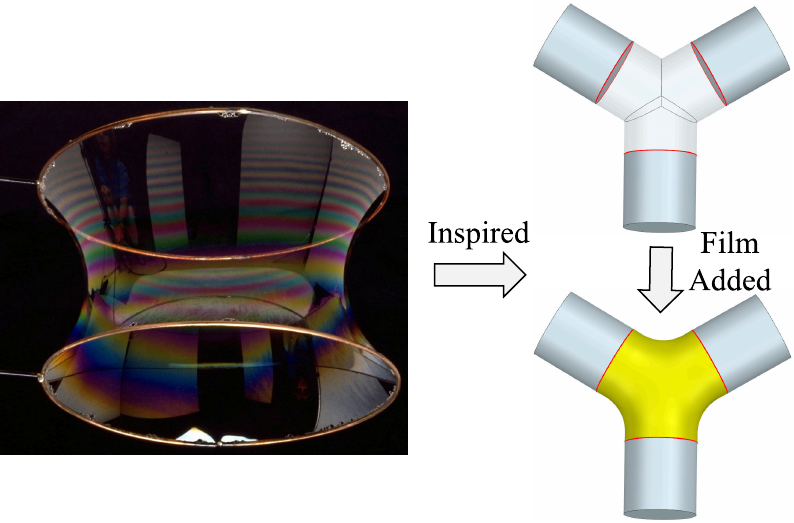}
\caption{The soap film-inspired nodal shape. (Soap film image source: https://www.quantamagazine.org/mathematicians-prove-melting-ice-stays-smooth-20211006/)}
\label{fig:soap}
\end{figure}

A soap film is a thin layer of liquid, typically composed of a soap-water solution, that is stabilized by surface tension. If two rings are dipped into a soapy solution and then pulled apart from an initially coincident state, a soap film naturally forms between them. As the film is stretched between the two rings, surface tension pulls the liquid into a stable configuration that minimizes the total surface area. Such a minimization process leads to a surface with zero mean curvature, ensuring a smooth transition between the two rings. A representative example of such a soap film is shown in Fig.~\ref{fig:soap}.

The behavior of soap film can be mathematically described using the Laplace equation~\cite{polygon-mesh-process}. This equation governs the equilibrium state of the surface by ensuring that the mean curvature remains zero, corresponding to an energy-minimizing configuration. It is given as:
\begin{equation}
    \nabla^2f = 0
\end{equation}
where the function $f$ represents a scalar field that depends on spatial coordinates, $\nabla^2$ (the Laplacian operator) represents the sum of second partial derivatives.

When working with discrete representations, such as polygonal meshes involved in this paper, the discrete Laplacian is used and given as:
\begin{equation}
\label{eq:laplacian}
    LV=0
\end{equation}
where $L$ is the Laplacian matrix of a mesh, and $V$ is a set of mesh vertex positions. By solving this linear system, a mesh can be faired while preserving the overall shape~\cite{1999_implicit-fairing}, leading to a smooth transition between the boundaries.

\begin{figure*}[htbp]
\centering
\includegraphics[width=\linewidth]{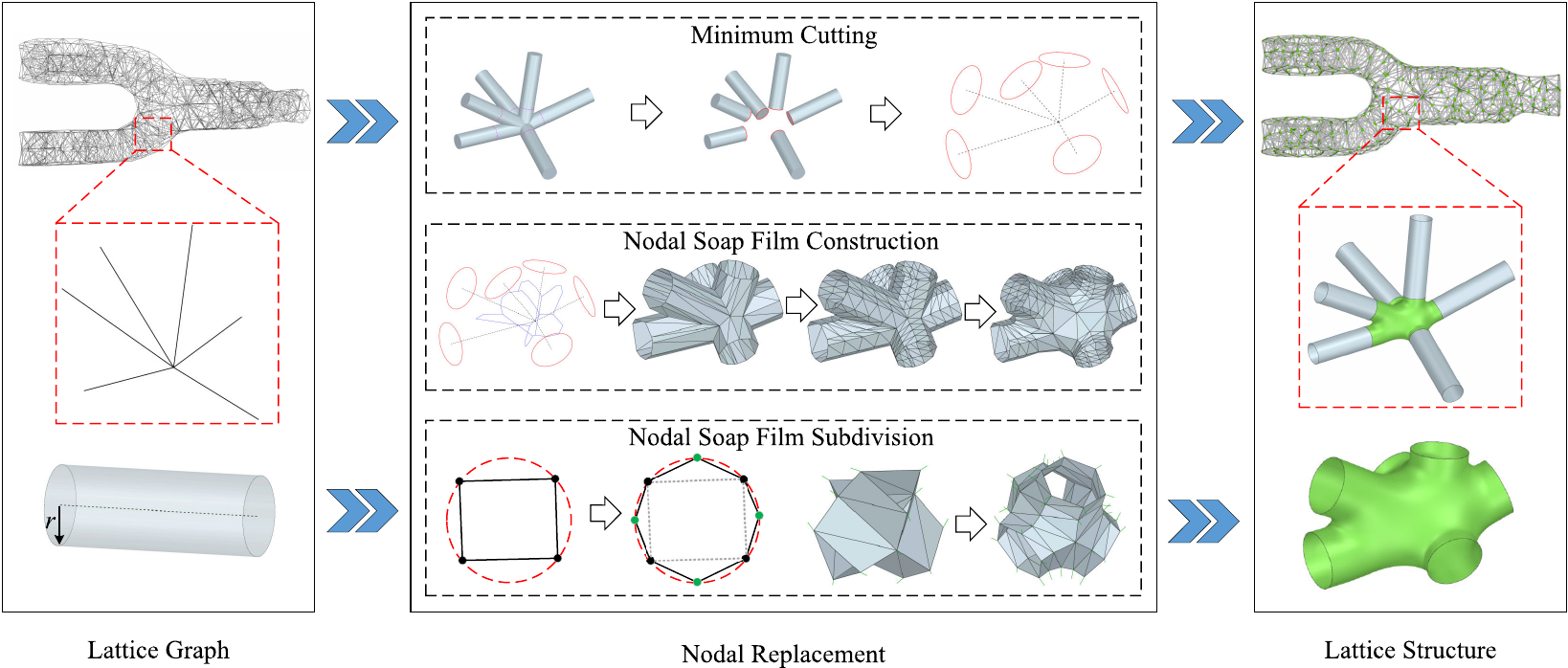}
\caption{The soap film-inspired framework.}
\label{fig:workflow}
\end{figure*}

Inspired by the behavior of soap films, we propose a novel framework to construct subdivisional nodal shapes to replace the original ones of a lattice structure, effectively addressing the challenge of managing surface-surface intersections. These subdivisional nodal shapes not only have reduced shape deviations and enhanced surface fairness but also ensure smooth transitions between nodal shapes and retained struts. A detailed overview of the framework is illustrated in Fig.~\ref{fig:workflow}.

 The framework begins with the input of a lattice graph, consisting of nodes, edges, and associated geometric parameters (e.g., cylinder radii assigned to edges), which collectively define the connectivity and strut geometry of the lattice structure. Subsequently, nodal replacement is implemented using a soap film-inspired nodal shape construction method. In the final stage, these newly generated nodal shapes are reassembled with the retained struts, forming a coherent lattice structure. As the key component of the framework, nodal replacement can be systematically divided into three main steps, as detailed below:
 
\textbf{Minimum cutting.} This step determines the cut positions for each strut to make the struts at a shared node not locally intersect after cutting while ensuring that the resulting void volume shrinks to a minimum. The resulting end circles of the cut struts then serve as boundary constraints for constructing the nodal soap films.

 \textbf{Nodal soap film construction.} This step generates the film geometry--the control mesh to be subdivided--for the replaceable portion of the lattice structure. The film geometry preserves the original node’s shape to minimize deviations between the reconstructed nodal shape and the original one. It also guarantees a smoother surface like soap film through Laplacian fairing.

 \textbf{Nodal soap film subdivision.} This step subdivides the constructed nodal soap film to generate a smooth and fair surface that approximates the original nodal shape. This subdivision process ensures smooth transitions between the nodal shapes and retained struts, leading to a smooth integration of them to form the overall lattice structure. 

 This work focuses on lattice structures with uniform cylindrical struts. Structures of this kind are among the most commonly used ones due to their simplicity and ease of manufacturing~\cite{2024_meta-meshing}.

\subsection{Minimum Cutting}
To construct subdivisional nodal shapes that replace the original ones, it is crucial to first cut the struts to eliminate the intersections between struts. However, excessive cutting can increase the void volume that is to be filled, potentially leading to significant shape modifications. Therefore, the cut position of each strut needs to be carefully determined to get a minimum cutting length, ensuring that, after cutting, the struts remain intersection-free while being retained as much as possible.

To illustrate the cutting strategy, we begin by considering the simple case involving only two struts, where the angle between their axes may be either less than or greater than 90 degrees. Subsequently, we extend the discussion to general cases involving multiple struts intersecting at a single node. The cutting process for these more complex scenarios can be viewed as the superposition of multiple simpler calculations, each corresponding to a pair of intersecting struts.

\begin{figure}[t]
\centering
\includegraphics[width=0.8\linewidth]{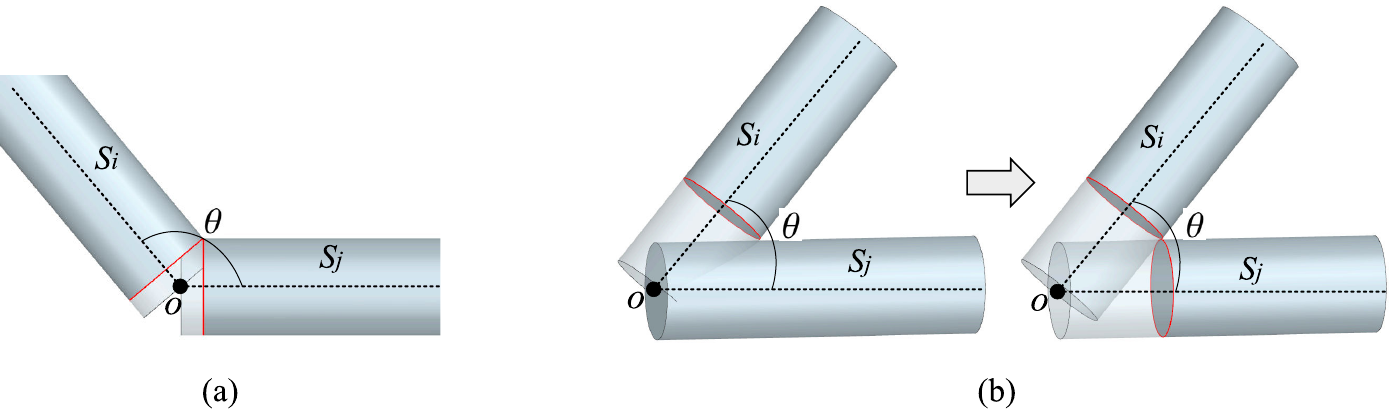}
\caption{The intersection-free cutting between two struts.}
\label{fig:intersection-free-cut}
\end{figure}

The cutting length computation of two cylinders intersecting at node $o$ involves struts $S_i$ and $S_j$, each with radius $r$, and the angle $\theta$ between their axes, as illustrated in Fig.~\ref{fig:intersection-free-cut}. If $\theta>90^\circ$, both struts have to be cut to achieve intersection-free contact and have the same minimum cutting length. In cases where $\theta<90^\circ$, achieving intersection-free contact initially requires cutting only one strut. However, such cutting may lead to intersections between the constructed nodal soap film and the retained struts. The nodal soap film construction requires that each swept cylinder—formed by sweeping an end circle along the strut axis toward the node—must not intersect with the retained struts (see detail construction process in Section~\ref{sec:film-geometry}). As a result, both struts need to be cut to the same length, and the minimum cutting length is given by:

\begin{equation}
\label{eq:collision-free}
    d_m=\frac{r}{tan(\theta/2)}
\end{equation}

When it comes to the general cases, for each strut $S_i$ at a given node, the above pair-wise minimum cutting is applied to all the other struts $S_j$. This process generates multiple candidate cutting lengths $d_{ij}$ for strut $S_i$, from which the final minimum cutting position $d^i_m$ is determined by selecting the maximum length. All the struts have intersection-free contacts and a minimized amount of cutting at the cutting length given by:
\begin{equation}
    d^i_m=\max d_{ij}, j=1,...,n \text{ and } j \neq i
\end{equation}

\begin{figure}[t]
\centering
\includegraphics[width=0.8\linewidth]{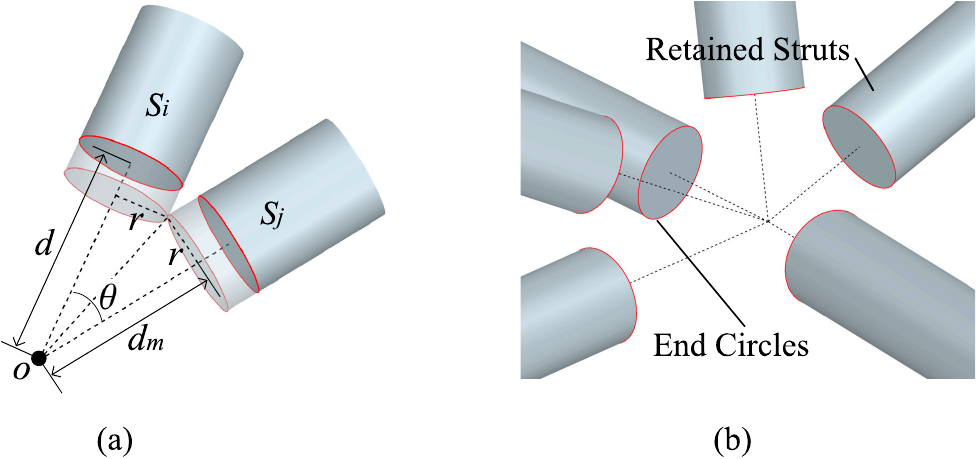}
\caption{The scaled cutting (a) and the cutting result of a general case (b).}
\label{fig:cutting}
\end{figure}

To prevent tangential contact between the end circles, which may result in degenerate cases such as overlapping facets in the constructed nodal soap film, the cutting length is slightly scaled (see Fig.~\ref{fig:cutting}a). The final cutting length $d^i$ for strut $S_i$ can be obtained by 
\begin{equation}
    d^i=(1+\lambda)d^i_m, i=1,...,n \text{ and }0<\lambda<0.5
\end{equation}

\subsection{Nodal Soap Film Construction}
\label{sec:film-geometry}
To construct the nodal soap film from given boundaries, the film geometry is first constructed by building a spherical Voronoi diagram using boundaries, where the diagram is adjusted to preserve the original node's shape. Then, this film geometry is smoothed to form a soap film-like surface.

\subsubsection{Film Geometry Construction}

\begin{figure}[t]
\centering
\includegraphics[width=0.8\linewidth]{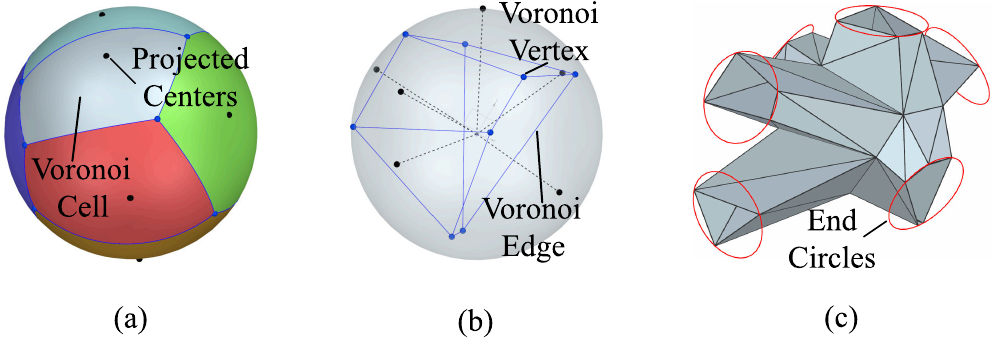}
\caption{The film geometry construction: (a) the Voronoi diagram on a sphere; (b) the Voronoi diagram with straight edges; and (c) the formed film geometry through projection.}
\label{fig:voronoi}
\end{figure}

The cutting algorithm leads to partial struts remaining and some end circles surrounding a void region, as shown in Fig.~\ref{fig:cutting}b. These circles are used as boundary constraints to construct the film geometry, that is, a control mesh to be subdivided to interpolate these boundary circles and fill the void region. To construct a topology-guaranteed film geometry in the void, a Voronoi-based approach is used, where the center points of the end circles are projected onto a sphere centered at the node, and then a spherical Voronoi diagram is generated using Fortune's algorithm~\cite{2011_spherical-fortune}. Such a diagram divides the sphere into multiple regions called Voronoi cells, making each cell correspond to an end circle, as depicted in Fig.~\ref{fig:voronoi}a. To form a triangular mesh based on this spherical Voronoi diagram, the arc edges on the sphere are straightened, forming polygonal cells, and the Voronoi vertices of each polygonal cell are projected onto their respective circles. The film geometry then can be constructed by connecting these Voronoi vertices to their corresponding projection points, as illustrated in Fig.~\ref{fig:voronoi}c.

To reduce deviations between the subdivisional and original nodal shapes, the film geometry needs to be closely aligned with the original node's shape, as the final shape of a subdivision surface is largely influenced by its initial control mesh. Achieving a high similarity between the film geometry and the original nodal shape can effectively reduce shape deviations of the reconstructed nodal geometry. For this purpose, adjustments have been made to the Voronoi diagram, including updating the positions of Voronoi vertices and inserting new vertices along the Voronoi edges. These modifications help approximate the intersection curves among cylinders, thereby improving the shape similarity of the film geometry.

\begin{figure*}[t]
\centering
\includegraphics[width=0.85\linewidth]{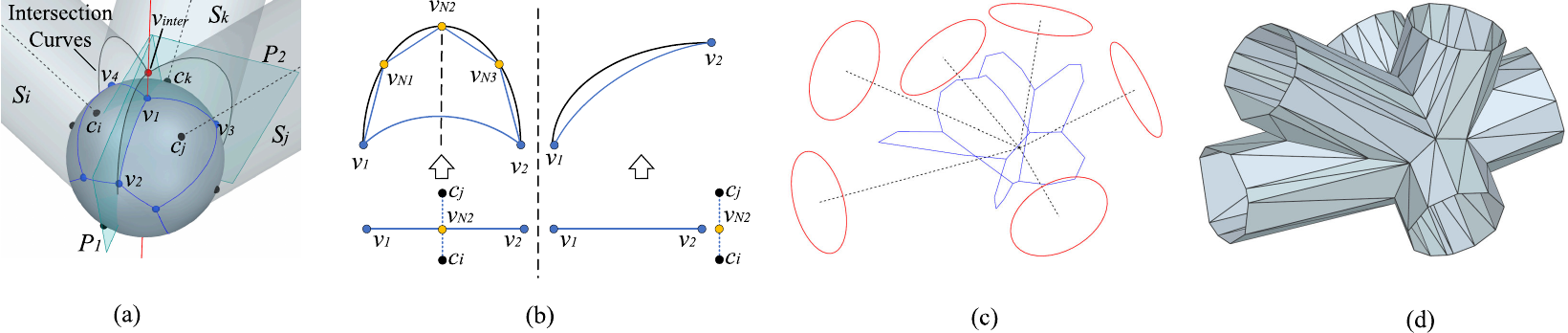}
\caption{Illustration of updating Voronoi diagram: (a) adjustment of Voronoi vertex positions; (b) insertion of new vertices along Voronoi edges; (c) the modified Voronoi diagram; and (d) the adjusted film geometry.}
\label{fig:adjustment}
\end{figure*}

The positions of existing Voronoi vertices are updated to coincide with the common intersection point of three struts. This intersection point can be determined using line-cylinder intersections rather than complicated surface-surface and surface-curve intersections, as both the Voronoi vertex and the intersection point lie on the same line. Taking Fig.~\ref{fig:adjustment}a as an illustration example, the projected center points $c_i$, $c_j$, and $c_k$ correspond to the three struts $S_i$, $S_j$, and $S_k$, while $v_1, v_2, v_3, v_4$ are Voronoi vertices on the sphere. It is known that the arcs $\overset{\frown}{v_1{v_2}}$ and $\overset{\frown}{v_1{v_3}}$ are equidistant from $c_i$ and $c_j$, and from $c_j$ and $c_k$, respectively. Two planes, $P_1$ and $P_2$, passing through two arcs and the node $o$, intersect each other, defining the line on which the Voronoi vertex $v_1$ is located. Similarly, the intersection curves of the corresponding cylinders $S_i$, $S_j$, and $S_k$ lie on these planes, ensuring that their common intersection point $v_{inter}$ also lies on the same line. Consequently, the updated position of the Voronoi vertex is obtained by calculating the intersection of the cylinder with the line oriented along the direction of $\overrightarrow{ov_1}$.

To approximate the intersection curves, a point insertion strategy is used, as illustrated in Fig.~\ref{fig:adjustment}b. This strategy begins by determining whether the intersection curves are monotonic using angle comparisons. If a curve is found to be non-monotonic, three new vertices are added, with their positions calculated via line-cylinder intersection computations. Specifically, the vector $\overrightarrow{ov_{N2}}$ is computed as $\overrightarrow{oc_i}+\overrightarrow{oc_j}$, $\theta=\measuredangle(\overrightarrow{ov_1}, \overrightarrow{ov_2})$, $\theta_1=\measuredangle(\overrightarrow{ov_1}, \overrightarrow{ov_{N2}})$, and $\theta_2=\measuredangle(\overrightarrow{ov_2}, \overrightarrow{ov_{N2}})$. If both  $\theta_1<\theta$ and $\theta_2<\theta$, the curve is classified as non-monotonic, which means the vector $\overrightarrow{ov_{N2}}$ can be projected onto the arc $\overset{\frown}{v_1{v_2}}$. The vertex $v_{N2}$ is then computed as the intersection of the line passing through $\overrightarrow{ov_{N2}}$ with the cylinder defined by the central axis $\overrightarrow{oc_i}$. The other two vertices $v_{N1}$ and $v_{N3}$ are generated by interpolating between $v_1$ and $v_{N2}$, and between $v_2$ and $v_{N2}$, respectively. Their positions are also refined through line-cylinder intersection computations. The updated Voronoi diagram and film geometry are shown in Fig.~\ref{fig:adjustment}c and d.

\subsubsection{Film Geometry Smoothing}
Although the shape similarity of the film geometry has been improved, the resulting shape directly subdivided from the film geometry may have poor overall surface fairness, stemming from the insufficient fairness of the film geometry at the transition regions between struts. Besides, such a subdivided nodal shape leads to non-smooth integration with retained struts at the boundaries where no smooth transition constraints are imposed.

To address these problems, a finer control mesh needs to be constructed from the film geometry to provide greater flexibility in redistributing vertex positions. This flexibility is required by satisfying two requirements. First, to increase fairness, the mesh needs to be smoothed while minimizing shape modification, leading to the need for local fairing. Second, to support smooth transitions, at least the 1-ring neighbors of boundary vertices need to remain constrained to the cylinder (see Section~\ref{sec:subdivision} for details).

\begin{figure}[t]
\centering
\includegraphics[width=0.8\linewidth]{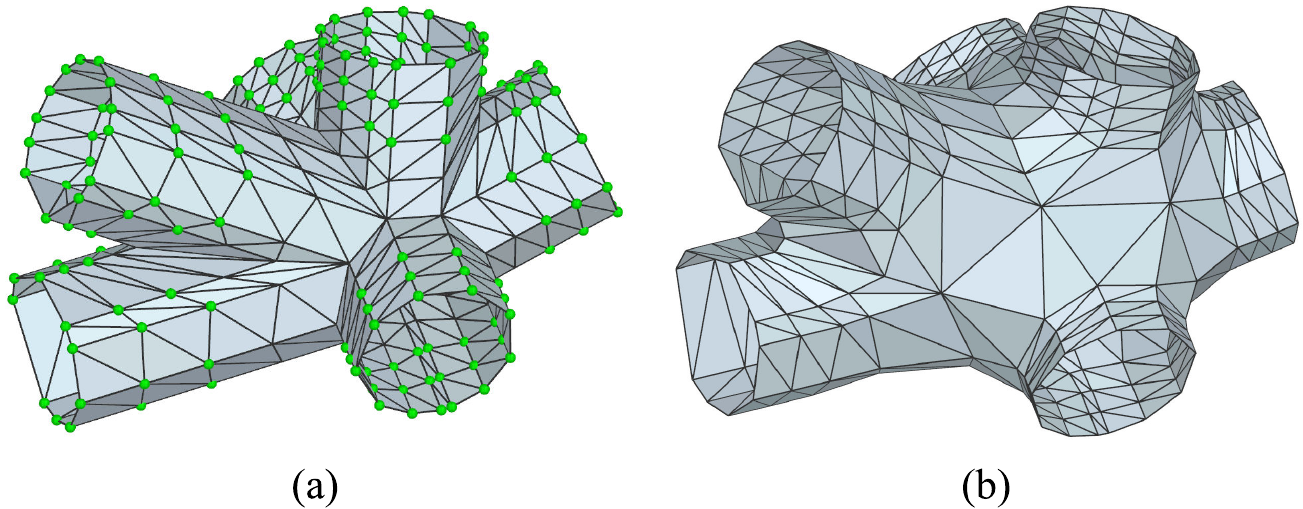}
\caption{The upsampled mesh with fixed vertices in green (a) and the smoothed control mesh (b).}
\label{fig:fairing}
\end{figure}

The finer mesh is generated through an upsampling process from the film geometry. Specifically, to balance the flexibility and vertex number, three layers of new vertices are inserted linearly along the mesh edges oriented in the direction of the cylinder axes. To ensure the new vertices conform to the cylindrical geometry, their positions are adjusted by projecting them perpendicular to the cylinder axes onto the cylinder surfaces. When performing smoothing, the 2-ring neighbors of boundary vertices (represented as small green spheres in Fig.\ref{fig:fairing}a) are kept fixed to maintain geometric constraints for smooth transition, while the remaining vertices are redistributed to improve surface fairness through solving the discrete Laplacian equation. To get a more visually appealing nodal shape (see Fig.~\ref{fig:fairing}b), it is smoothed using a second-order Laplacian matrix~\cite{1999_implicit-fairing}.

\subsection{Nodal Soap Film Subdivision}
\label{sec:subdivision}
To generate subdivisional nodal shapes that integrate smoothly with retained struts, a combined PN-Loop scheme is employed to subdivide the nodal soap film. This scheme utilizes circle sampling for boundary subdivision, ensuring that the limit surface precisely interpolates the boundary curves. The PN-Loop scheme is then applied to subdivide the inner mesh, producing a smooth and fair surface while maintaining smooth transitions at the boundaries.

\subsubsection{Boundary Subdivision}
\begin{figure}[t]
\centering
\includegraphics[width=0.8\linewidth]{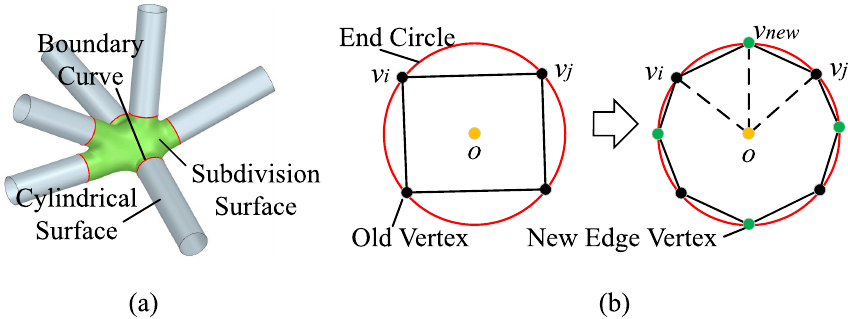}
\caption{The boundary subdivision: (a) shared boundary curves; (b) the circle sampling for boundary refinement.}
\label{fig:boundary}
\end{figure}

With the proposed modeling method, the final surface of the lattice structure is the union of cylindrical surfaces and subdivision surfaces, as illustrated in Fig.~\ref{fig:boundary}a. To ensure geometric continuity, the cylindrical and subdivision surfaces must share common boundary curves without gaps. This requirement necessitates that the subdivision surface interpolates the circular boundaries. Consequently, specialized subdivision rules should be applied at the boundaries, which consider boundary conditions to calculate the points on the next refinement level.

For the boundary refinement, we use a simple rule called circle sampling, as depicted in Fig~\ref{fig:boundary}b. Assuming that the boundary circle is given in the form of a parametric curve $c(u)$ and the boundary vertices of the initial control mesh, generated via Voronoi vertex projection, lie on the given curve and form a closed loop. For any boundary edge in this loop, defined by two vertices $v_i$ and $v_j$ with corresponding parametric values $u_i$ and $u_j$, a new edge vertex $v_{new}$ is calculated by the appropriate sampling of the given curve.
\begin{equation}
    v_{new}=c(\frac{u_i+u_j}{2})
\end{equation}

Meanwhile, the original boundary vertices are directly retained as new vertex vertices during subdivision. By explicitly incorporating the given boundary curves in each iteration, this rule ensures that the limit surface satisfies the boundary interpolation conditions, facilitating the seamless integration of nodal shapes and retained struts.

\subsubsection{PN-Loop Subdivision} 
\begin{figure}[t]
\centering
\includegraphics[width=0.8\linewidth]{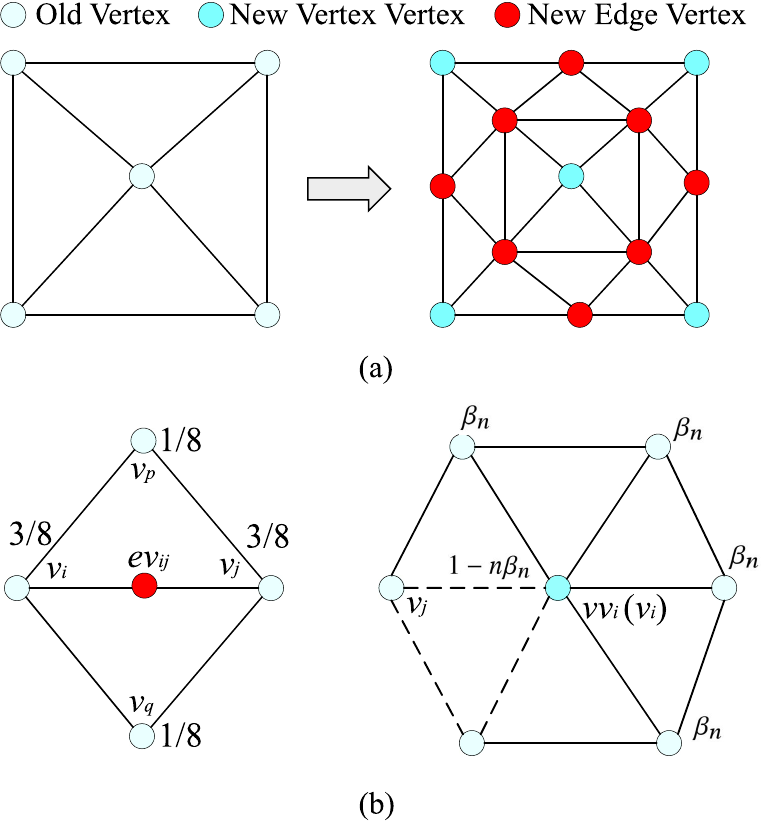}
\caption{The Loop subdivision scheme: (a) the topological rules; (b) the geometric rules.}
\label{fig:loop}
\end{figure}

At the interior of the control mesh, we apply the PN-Loop subdivision scheme to refine the control mesh. The PN-Loop scheme is derived by instantiating the generalized point-normal subdivision framework proposed by Yang~\cite{2023_PN-subdivision} on the classical Loop subdivision scheme~\cite{loop1987smooth}. The Loop scheme iteratively refines a mesh by splitting the triangles into four smaller triangles, with the positions of vertices being updated through averaging, as illustrated in Fig.~\ref{fig:loop}. Specifically, for the new vertex vertex $vv_i$ corresponding to $v_i$, its positions are updated by
\begin{equation}
    vv_i=(1-n\beta_n)v_i + \beta_n\sum_{j=0}^{n-1}v_j 
\end{equation}

\begin{equation}
    \beta_n=\frac{1}{n}(\frac{5}{8}-(\frac{3}{8}+\frac{1}{4}cos\frac{2\pi}{n})^2)
\end{equation}
where $n$ indicates the valence of vertex $v_i$, and $v_j$ is the neighborhood vertex of $v_i$ within 1 ring.

The positions of the new edge vertex $ev_{ij}$ corresponding to the edge $v_iv_j$ are computed by
\begin{equation}
    ev_{ij}=\frac{3}{8}(v_i+v_j)+\frac{1}{8}(v_p+v_q)
\end{equation}

By generalizing the Loop scheme, making it refine polygon meshes with the refinement of unit control normals at the vertices, the PN-Loop scheme can be obtained. Let the initial control points and corresponding unit control normals of a triangular mesh be denoted as ${(v_i^0, n_i^0): i=1,...,n}$. The PN-Loop subdivision refines the control mesh with initial control normals through the following iterative computations. 

For new vertex vertex:
\begin{align}
\left\{
\begin{aligned}
t_i^{k+1}&=(1-n\beta_n)v_i^k + \beta_n\sum_{j=0}^{n-1}v_j^k \\
n_i^{k+1}&=\frac{(1-n\beta_n)n_i^k + \beta_n\sum_{j=0}^{n-1}n_j^k}{\|(1-n\beta_n)n_i^k + \beta_n\sum_{j=0}^{n-1}n_j^k\|} \\
h_i^{k+1}&=(1-n\beta_n)\frac{(n_i^k+n_i^{k+1})^{\top}(v_i^k-t_i^{k+1})}{(n_i^k+n_i^{k+1})^{\top}n_i^{k+1}} \\
         &+\beta_n\sum_{j=0}^{n-1}\frac{(n_j^k+n_i^{k+1})^{\top}(v_j^k-t_i^{k+1})}{(n_j^k+n_i^{k+1})^{\top}n_i^{k+1}} \\
vv_i^{k+1}&=t_i^{k+1}+h_i^{k+1}n_i^{k+1}
\end{aligned}
\right.
\end{align}

For new edge vertex:
\begin{align}
\left\{
\begin{aligned}
t_{ij}^{k+1}&=\frac{3}{8}(v_i^k+v_j^k)+\frac{1}{8}(v_p^k+v_q^k)\\
n_{ij}^{k+1}&=\frac{\frac{3}{8}(n_i^k+n_j^k)+\frac{1}{8}(n_p^k+n_q^k)}{\|\frac{3}{8}(n_i^k+n_j^k)+\frac{1}{8}(n_p^k+n_q^k)\|}\\
h_{ij}^{k+1}&=\frac{3}{8}(\frac{(n_i^k+n_{ij}^{k+1})^{\top}(v_i^k-t_{ij}^{k+1})}{(n_i^k+n_{ij}^{k+1})^{\top}n_{ij}^{k+1}}+\frac{(n_j^k+n_{ij}^{k+1})^{\top}(v_j^k-t_{ij}^{k+1})}{(n_j^k+n_{ij}^{k+1})^{\top}n_{ij}^{k+1}})\\
&+\frac{1}{8}(\frac{(n_p^k+n_{ij}^{k+1})^{\top}(v_p^k-t_{ij}^{k+1})}{(n_p^k+n_{ij}^{k+1})^{\top}n_{ij}^{k+1}}+\frac{(n_q^k+n_{ij}^{k+1})^{\top}(v_q^k-t_{ij}^{k+1})}{(n_q^k+n_{ij}^{k+1})^{\top}n_{ij}^{k+1}})\\
ev_{ij}^{k+1}&=t_{ij}^{k+1}+h_{ij}^{k+1}n_{ij}^{k+1}
\end{aligned}
\right.
\end{align}

\begin{figure}[t]
\centering
\includegraphics[width=0.8\linewidth]{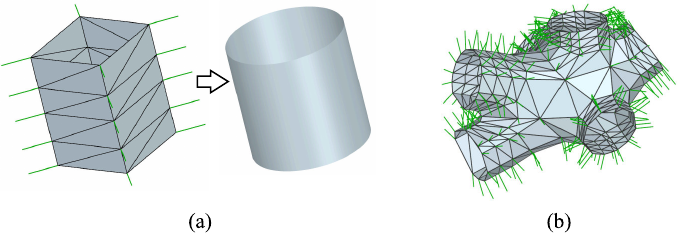}
\caption{The cylinder-preserving feature (a) and the normal configuration of combined PN-Loop subdivision scheme (b).}
\label{fig:PN-Loop}
\end{figure}

This scheme has the same $C^1$ smoothness as the Loop scheme (the relationship between the smoothness of the generalized PN scheme and that of the traditional scheme has been thoroughly examined in~\cite{2023_PN-subdivision}), ensuring the generation of a smooth and fair surface from the constructed nodal soap film. It is notable that the PN-Loop scheme introduces a unique capability of reproducing cylinders, which ensures that when the initial control points and normals are sampled from cylindrical geometries, the resulting subdivision surface can accurately capture the geometric shape, as demonstrated in Fig.~\ref{fig:PN-Loop}a. This cylinder preservation property has been analyzed in detail by Yang~\cite{2023_PN-subdivision} and plays a critical role in achieving smooth transitions at the boundaries in our method.

\begin{figure*}[t]
\centering
\includegraphics[width=0.85\linewidth]{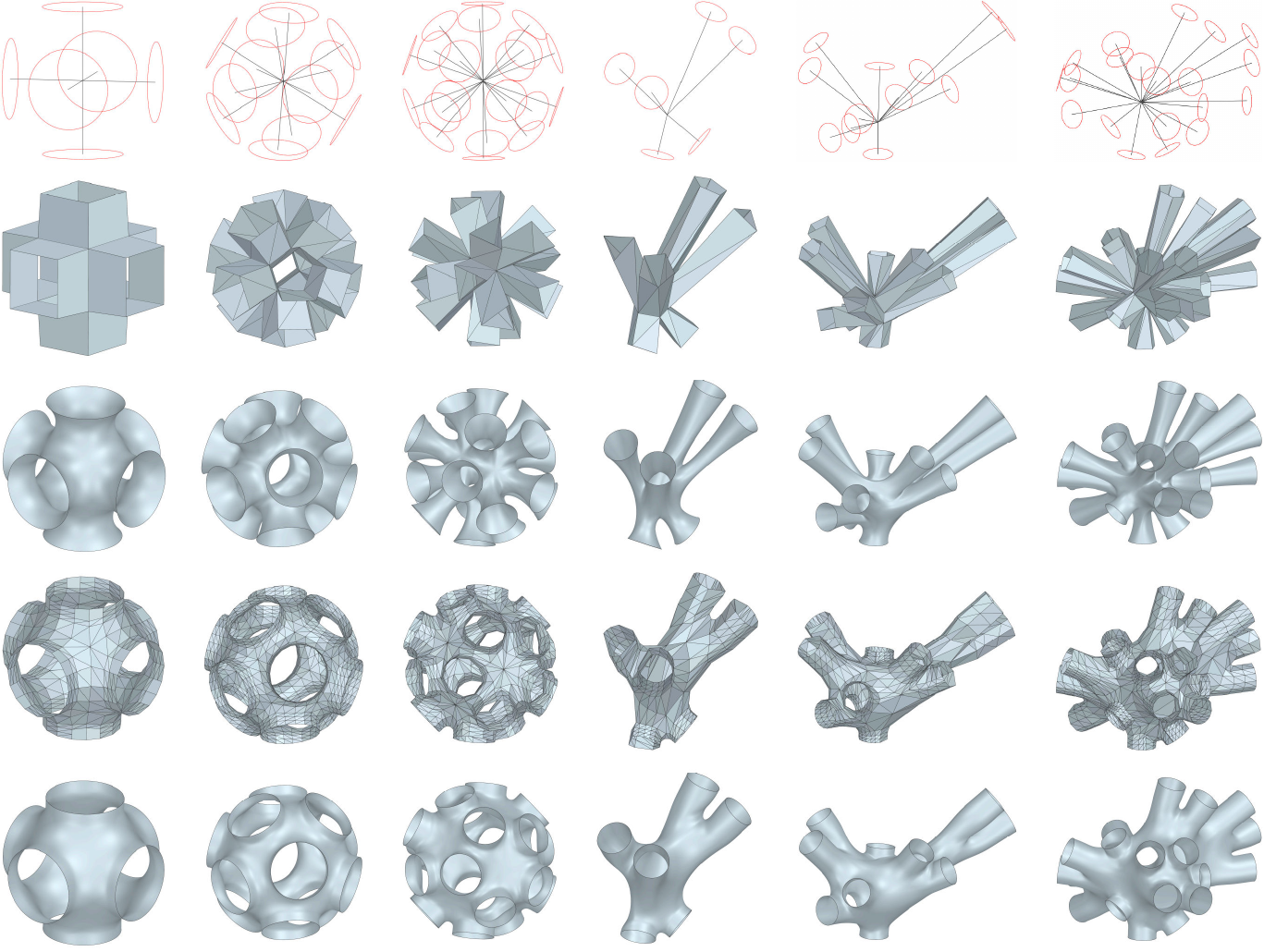}
\caption{The representative nodes (the top row) and their resulting control meshes and subdivision surfaces generated by Savio's method (the second and third rows) and our method (the below two rows).}
\label{fig:nodes}
\end{figure*}

\begin{figure}[t]
\centering
\includegraphics[width=0.8\linewidth]{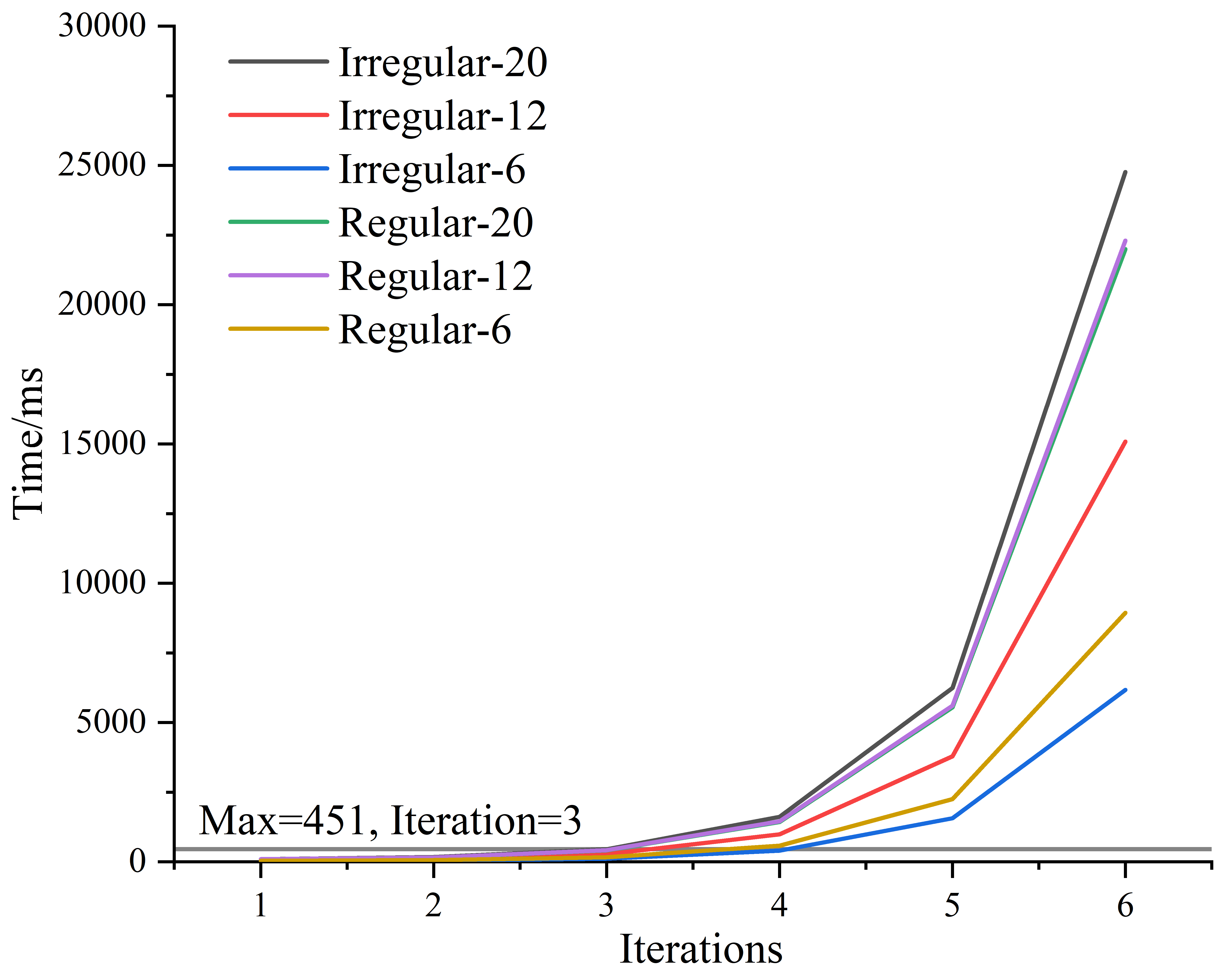}
\caption{The relationship between subdivision time and iterations.}
\label{fig:iter-time}
\end{figure}

To ensure smooth transitions between the subdivisional nodal shapes and the retained struts, particular emphasis needs to be placed on configuring vertex normals. For boundary vertices and their 1-ring neighbors, their normals are explicitly assigned with the external cylinder normals. These point-normal pairs make the newly subdivided points and normals by PN subdivision involving them lie on the same cylinder. As a result, the PN subdivision surface converges to the cylinder in the boundary region. Thus, the retained struts and subdivision surfaces interpolate the same boundary curves and have the same cylindrical shape near boundaries, naturally achieving smooth transitions. For all other vertices, normals are computed by aggregating the face normals of the adjacent triangles and normalizing the result~\cite{2016_vertex-normal-estimation}. Such a tailored normal configuration is illustrated in Fig.~\ref{fig:PN-Loop}b.

After completing the nodal replacements, the subdivisional nodal shapes fill the void regions, resulting in a B-rep lattice structure. This B-rep model consists of only three types of boundary elements: cylindrical surfaces defined by pairs of circles, subdivision surfaces constructed from circles surrounding individual nodes, and boundary curves corresponding to the end circles of the cylinders. Additionally, the quantities of these elements can be directly determined from the lattice graph's topology. Specifically, for a lattice graph with $N_e$ edges and $N_n$ nodes, the B-rep model includes $N_e$ cylindrical surfaces, $N_n$ subdivision surfaces, and $2N_e$ boundary curves.

\begin{figure*}[t]
\centering
\includegraphics[width=0.9\linewidth]{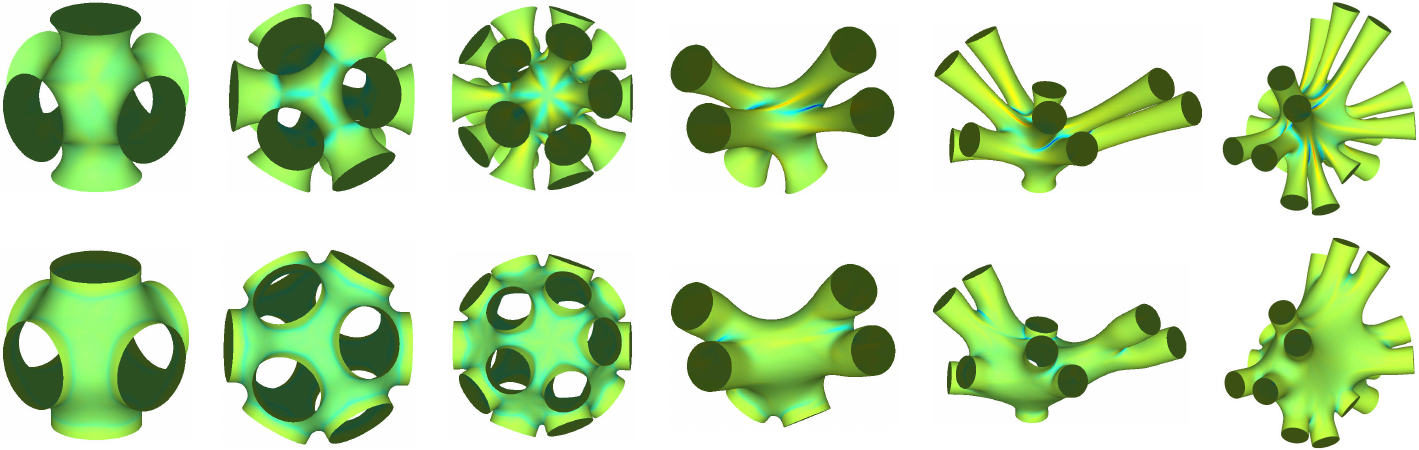}
\caption{Mean curvatures of the subdivision surfaces generated by Savio's method (the top row) and our method (the below row), changing from (positive) high values through zero to (negative) low values when the colors change from red through green to blue.}
\label{fig:smoothness}
\end{figure*}

\begin{figure*}[t]
\centering
\includegraphics[width=0.93\linewidth]{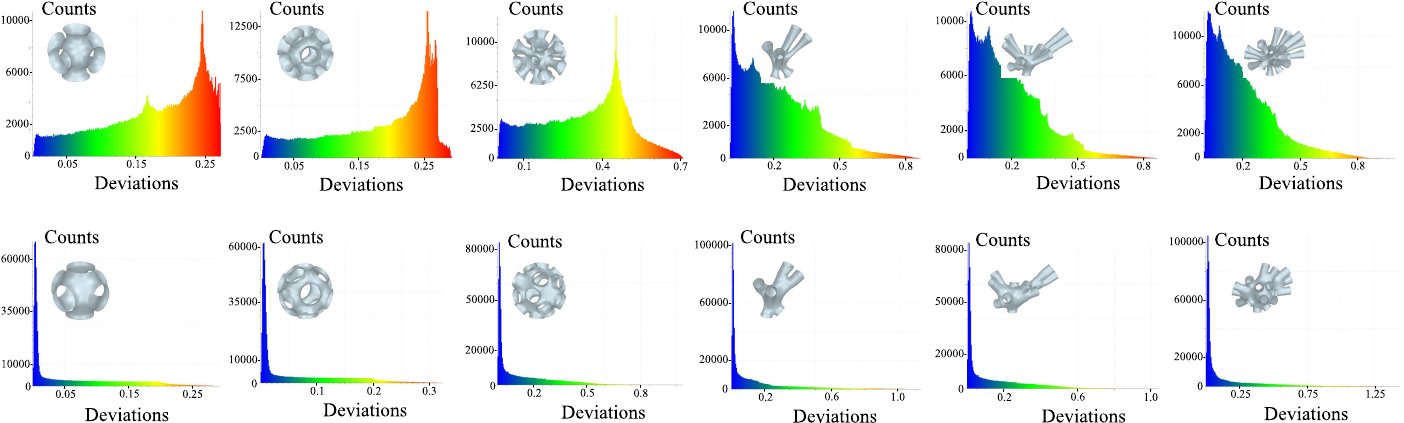}
\caption{The distributions of point deviations, changing from low values to high values when colors change from blue through green to red.}
\label{fig:deviation}
\end{figure*}

\begin{table*}[t]
\caption{Comparison of the proposed method with Savio's method~\cite{2018_subdivision_2Dsurface}}
    \centering
    \resizebox{\linewidth}{!}{%
    \begin{tabular}{c c c c c c}
    \hline
    \multirow{2}*{Models}&\multirow{2}*{Methods}&\multicolumn{3}{c}{Deviations/mm}&{Improvements}\\
    \cline{3-5}
    & &Max&Avg&Std&(w.r.t. Avg Deviation)\\
    \hline
    \multirow{2}*{Regular-6}&Our method& 0.2846& 0.0647& 0.0715&\multirow{2}*{ 62.6\%}\\
    &Savio's method& 0.2832& 0.1732& 0.0737&\\
    \hline
    \multirow{2}*{Regular-12}&Our method& 0.3021& 0.0645& 0.0751&\multirow{2}*{ 64.3\%}\\
    &Savio's method& 0.3021& 0.1808& 0.0793&\\
    \hline
    \multirow{2}*{Regular-20}&Our method& 0.6653& 0.1009& 0.1394&\multirow{2}*{ 69.2\%}\\
    &Savio's method& 0.6958& 0.3279& 0.1666&\\
    \hline
    \multirow{2}*{Irregular-6}&Our method& 1.1082& 0.1643& 0.2205&\multirow{2}*{ 27.1\%}\\
    &Savio's method& 0.8457& 0.2254& 0.1659&\\
    \hline
     \multirow{2}*{Irregular-12}&Our method& 0.9762& 0.1442& 0.1690&\multirow{2}*{ 26.9\%}\\
    &Savio's method& 0.8029& 0.1972& 0.1481&\\
    \hline
     \multirow{2}*{Irregular-20}&Our method& 1.3798& 0.1681& 0.2480&\multirow{2}*{ 18.4\%}\\
    &Savio's method& 0.9722& 0.2059& 0.1601&\\
    \hline
    \end{tabular}}
    \label{tab:comparison-results}
\end{table*}

\begin{table*}[t]
\caption{The statistics related to lattice structure construction.}
    \centering
    \resizebox{\linewidth}{!}{%
    \begin{tabular}{c c c c c c c c c c}
    \hline
    \multirow{2}*{Models}&\multirow{2}*{NO. of Edges}&\multirow{2}*{NO. of Nodes}&\multirow{2}*{Max Degree}&\multicolumn{3}{c}{Smoothing Time/ms}&\multicolumn{3}{c}{Construction Time/ms}\\
    \cline{5-10}
    & & & &Min&Max&Avg&Min&Max&Avg\\
    \hline
     Heat Exchanger& 6125& 4908&39&5 &94 &29.5 &40 &649 &216.8 \\
    Lever Arm& 12960&2665 &38&6 &99 &28.6 &48 &677 &205.6 \\
     Mounting Bracket& 24685& 4873&38&6 &108 &31.7 &50 &761 &230.3 \\
    Surfacing Mold& 26813& 5306&33&7 &90 &31.7 &53 &633 &229.3 \\
    Engine Bracket& 33005& 6517&35&6 &94 &30.4 &48 &652 &218.6 \\
    Bevel Gear& 38325& 7602&40&6 &105 &31.1 &49 &727 &227.6 \\
    Pipe Holder& 46331& 9134&41&8 &105 &31.3 &60 &732 &230.7 \\
    Impeller& 47377& 9283&38&6 &107 &32.7 &50 &734 &236.6 \\
    Gear Box& 79200&15571 &38&5 &105 &32.5 &39 &707 &227.8 \\
    \hline
    \end{tabular}}
    \label{tab:time-results}
\end{table*}

\begin{figure*}[t]
\centering
\includegraphics[width=0.8\linewidth]{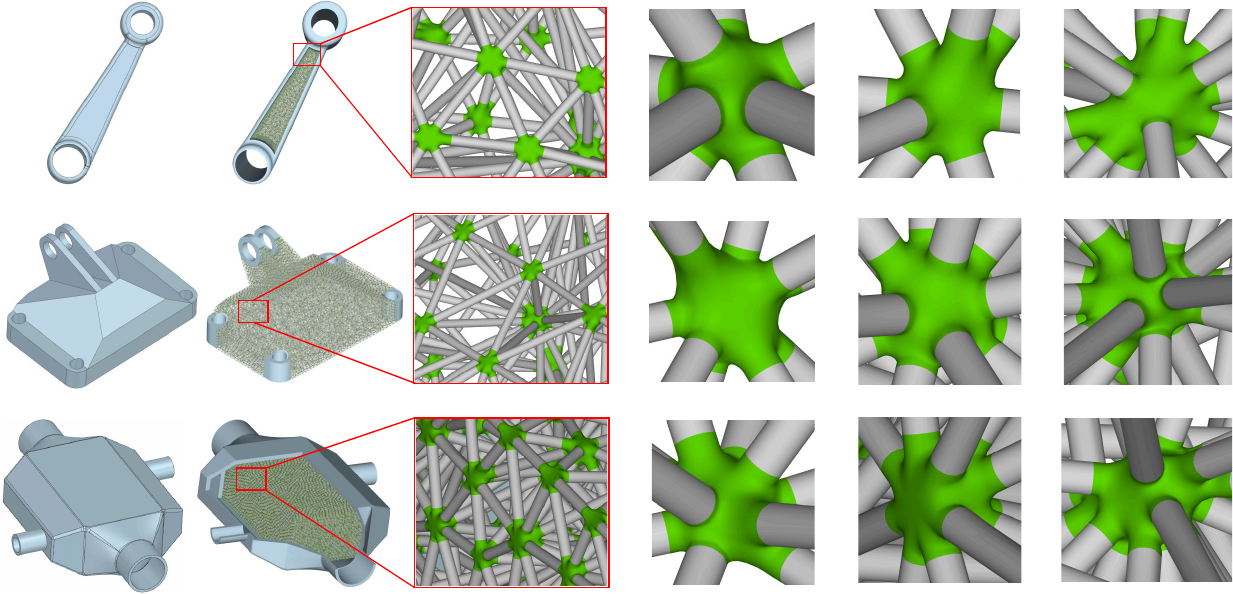}
\caption{The resulting subdivisional lattice structures of simple cases.}
\label{fig:simple}
\end{figure*}

\begin{figure*}[t]
\centering
\includegraphics[width=0.8\linewidth]{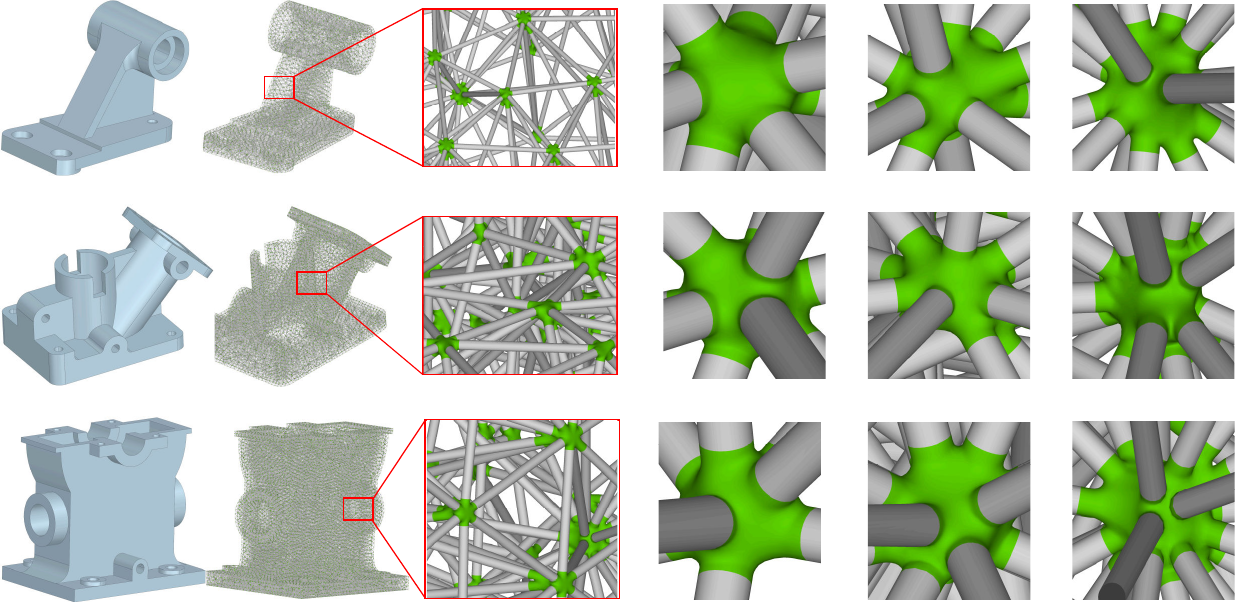}
\caption{The resulting subdivisional lattice structures of moderate cases.}
\label{fig:moderate}
\end{figure*}

\begin{figure*}[t]
\centering
\includegraphics[width=0.8\linewidth]{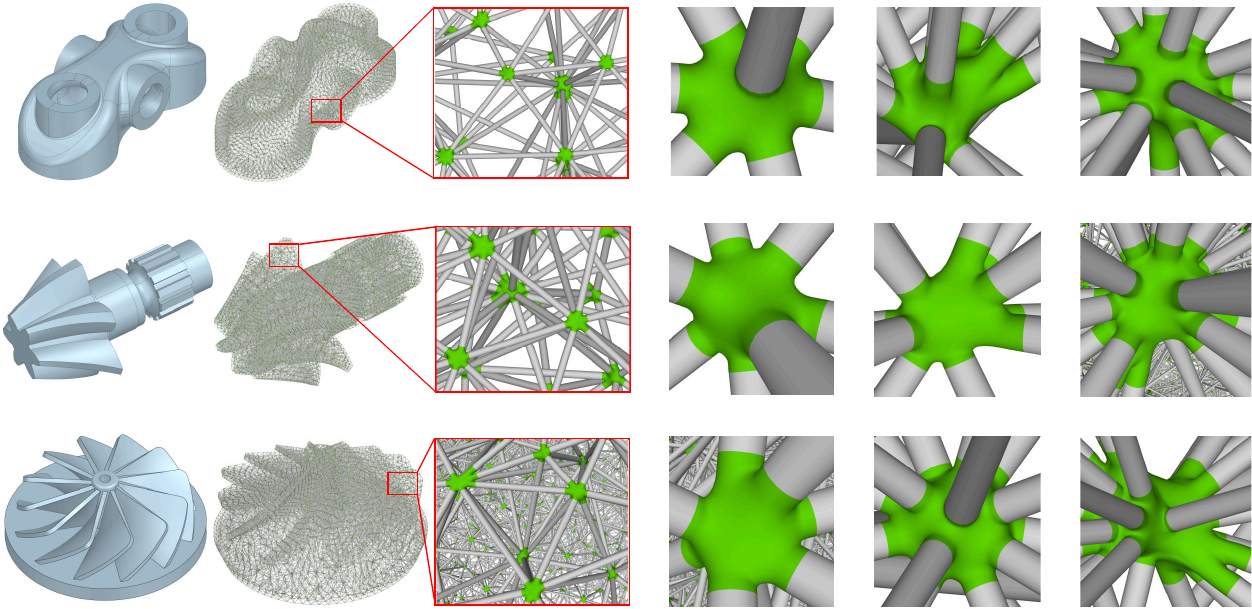}
\caption{The resulting subdivisional lattice structures of complex cases.}
\label{fig:complex}
\end{figure*}

\section{Results}
\label{sec:results}
The proposed method has been implemented in C++ on a system running Ubuntu 20.04, equipped with an Intel Core i9-12900K CPU operating at 5.20 GHz and 128 GB of RAM. All geometric computations were carried out using the Open CASCADE geometric modeling kernel (version 7.6.0) and the software framework OpenMesh (version 11.0), designed for representing and manipulating polygonal meshes. To evaluate the method's effectiveness, 10 case studies were conducted, with results demonstrating its capability across a variety of scenarios. Case Study 1 (Fig.\ref{fig:nodes}) considered 6 nodes, with their topology configurations ranging from regular to irregular and simple to complex. They were also used to compare the proposed method with Savio's approach~\cite{2018_subdivision_2Dsurface}. Case Studies 2–10 (Fig.~\ref{fig:simple}-\ref{fig:complex}) involved 9 lattice structures, which are classified into three categories based on the shape complexity of the original solid models.

\subsection{Examples}
Case study 1 considered six nodes (see the top row of Fig.~\ref{fig:nodes}), which are categorized into regular (the first three) and irregular (the last three) and each category includes nodes with 6, 12, and 20 edges. These nodes were selected as representative examples, encompassing a range of topological configurations from simple and regular to complex and irregular. When cutting these nodes, they are calculated assuming a cylinder radius of 1 mm and a length scaling parameter of $\lambda=0.3$. To evaluate the surface fairness and shape deviation of the nodal shapes generated by our method, we compare them with those obtained using Savio's method~\cite{2018_subdivision_2Dsurface}. 

We generated the control meshes and subdivision surfaces using our method and Savio's method~\footnote{In Savio's method, the whole lattice structure is modeled using subdivision surfaces without specific consideration for boundary curves. Because we just need to compare the nodal shapes, the boundary subdivisions were integrated into Savio's method to align the retained struts with those in our approach.}, respectively. The number of subdivision iterations for our method was set to three, based on an analysis of the relationship between subdivision time and the number of iterations, as shown in Fig.\ref{fig:iter-time}. This analysis revealed a sharp increase in time consumption beyond three iterations. In contrast, Savio's method employs five subdivision iterations due to the involvement of a relatively coarser control mesh. The comparative results are presented in Fig.\ref{fig:nodes}. Then, the mean curvatures of two kinds of subdivision surfaces are computed to measure the surface fairness, as shown in Fig.~\ref{fig:smoothness}. In addition, to accurately analyze the shape deviations of subdivisional nodal shapes, we sampled eight hundred thousand points for each shape and computed the closest distance to the original nodal shapes at each point, treating these distances as measures of shape deviation. Fig.~\ref{fig:deviation} shows the distributions of these deviations and the deviation statistics, including the maximum, average, and standard deviations, are summarized in Table~\ref{tab:comparison-results}.

Case studies 2-10 (Fig.~\ref{fig:simple}-\ref{fig:complex}) generated subdivision surface-based lattice structures from 9 lattice graphs, which were obtained using the TetGen tool from 9 solid models. Case studies 2–4 involve three relatively simple models (level arm, engine bracket, and heat exchanger), featuring simple shapes and surfaces. Case studies 8–10 involve complex geometries with freeform surfaces (surfacing mold, bevel gear, and impeller). Case studies 5–7 represent an intermediate level of complexity, combining moderately intricate shapes and surfaces (mounting bracket, pipe holder, and gear box). To prevent invalid cuts where the cutting length exceeds the strut length, a conservative strut radius is applied, empirically set to 1/8 of the shortest edge length in the lattice graph. The length scaling parameter $\lambda$ is also set to 0.3.

Among these, case studies 2 and 3 illustrate the generation of lattice structures in specific regions of mechanical components, preserving essential features such as cylindrical holes to achieve lightweight designs. Case study 4 demonstrates the application of lattice structures within a heat exchanger, enhancing heat dissipation efficiency. These examples underscore the practical utility of lattice structures in engineering applications, while the remaining cases demonstrate the effectiveness of the proposed method for subdivisional construction of lattice structures. Table~\ref{tab:time-results} presents a summary of some statistics related to lattice structure construction, including the number of lattice graph's edges, the max degree of nodes, the minimum, maximum, and average times required for film geometry smoothing, and subdivisional nodal shape construction.

\subsection{Discussions and Limitations}

For the six representative nodes considered above, the subdivisional nodal shapes have been constructed and compared, showing significant improvements in surface smoothness and shape similarity to the original nodal shapes. The lattice structures have also been accurately modeled from lattice graphs composed of various topologically complex nodal configurations, as expected. All these cases significantly demonstrate the versatility and effectiveness of our method for subdivisional lattice structure construction.

From the colored curvature image in Fig.~\ref{fig:smoothness}, we can see that the subdivision surfaces generated using our method exhibit superior surface fairness compared to those generated by Savio's method, particularly for irregular cases. In these cases, surfaces produced by Savio's method exhibit high mean curvatures in the red and blue regions, confirming the advance of constructing a smooth control mesh. In addition, the deviation stats in Table~\ref{tab:comparison-results} further confirm the advantages of the proposed method. All six cases show effective reductions in average shape deviations. It should be noted that in some cases, the maximum deviation in our method may exceed that of Savio's method. Our focus, however, is on the overall quality of the subdivisional nodal shapes, which is better reflected by the average deviation rather than the extreme value represented by the maximum deviation at a single point. This is supported by the distribution of point deviations for the subdivisional nodal shapes generated by our method in Fig.~\ref{fig:deviation}, where, despite the presence of large deviations, the majority of deviations remain small, resulting in a lower average deviation.

The statistics in Table~\ref{tab:time-results} highlight the computational efficiency and consistency of the proposed method. On average, it generates a subdivisional nodal shape in approximately 0.23 seconds, independent of the lattice graphs' complexity, demonstrating consistent performance across different models. Of this time, around 13\% (approximately 0.03 seconds) is allocated to smoothing the control mesh. This minor addition to the computation time is justified, as it ensures a smooth and fair surface. Generally, the time consumption primarily depends on the node degree, as higher-degree nodes generate a greater number of vertices and facets during nodal soap film construction, leading to higher computational demands for smoothing and subdivision.

\section{Conclusions}
\label{sec:conclusion}
A new method has been presented in this paper to construct lattice structures from lattice graphs, featuring a soap film-inspired framework to construct smooth and fair nodal shapes. This approach can effectively avoid complex surface-surface intersections to ensure robustness in the construction and minimize shape deviations from the original nodal geometry. The method achieves this by constructing a nodal soap film that approximates the original nodal geometry, smoothing the film geometry through Laplacian fairing, subdividing the nodal soap film using a combined PN-Loop subdivision scheme with enforced smooth transitions at boundaries, and assembling these nodal shapes with retained struts to yield a B-rep model. A series of case studies and comparisons have also been conducted to validate the method.

While the proposed method has demonstrated effectiveness in the case studies conducted, certain limitations should be acknowledged. For example, the smooth transition at boundaries relies on the PN scheme’s capability to reproduce cylindrical shapes, restricting the current approach to lattice structures with cylindrical struts. Although these types are widely used, other lattice structures, such as those with quadric profiles~\cite{2018_quador,2021_slicing_efficiency_matrix-oriented-data-structure} or curved axes~\cite{2021_1D_curve-axis_Bai_curved-lattice-strut,2021_1D_curve-axis_Fu_curved-beam}, have shown promising applications in various fields. Expanding the method to accommodate a broader range of lattice structures remains an important direction for future research.

Additionally, our method assumes valid nodal configurations, where conservative cutting lengths remain less than half of the edge lengths. However, when the angles between edges at a node are very small, larger cutting lengths may be obtained, leading to a nodal shape with significant shape deviations. In more extreme cases, the cutting length can exceed the edge length, resulting in invalid outputs. Extending the proposed method to address such scenarios or relaxing the assumption presents a compelling direction for future improvement.

Lastly, AI methods---particularly generative approaches---have demonstrated significant potential in areas such as natural language processing~\cite{achiam2023gpt} and computer vision~\cite{yang2023diffusion}, and are increasingly being applied to the field of geometric modeling~\cite{zou2024intelligent,zou2025splinegen}. The application of AI to the construction of lattice structures, especially in addressing the associated decision-making challenges, represents a promising direction for future research.

\section*{Acknowledgements}
This work has been funded by NSF of China (No. 62102355) and the ``Pioneer" and ``Leading Goose" R\&D Program of Zhejiang Province (No. 2024C01103).

 \bibliographystyle{elsarticle-num} 
 \bibliography{cas-refs}





\end{document}